\documentclass[aps,prd,11pt,preprint,superscriptaddress,nofootinbib,floatfix]{revtex4}

\usepackage{latexsym}
\usepackage{amsfonts}
\usepackage[latin1]{inputenc}
\usepackage[caption=false]{subfig}
\usepackage{graphicx}
\usepackage{mathrsfs}
\usepackage{amssymb}
\usepackage{amsmath}
\usepackage{verbatim}
\usepackage{multirow}
\usepackage{color}
\usepackage{epsfig}
\usepackage{amscd}
\usepackage{bm}
\usepackage{hyperref}

\def\D0{\slash\!\!\!\!\!\!\!\!\!\:D0}

\begin{document}

\preprint{PSI-PR-13-19}
\date{\today}

\title{Composite Higgs: searches for new physics at future $e^+e^-$ colliders}

\author{D.~Barducci}\email[E-mail: ]{d.barducci@soton.ac.uk}
\affiliation{School of Physics and Astronomy, University of Southampton, Southampton SO17 1BJ, U.K.}
\author{S.~De~Curtis}\email[E-mail: ]{decurtis@fi.infn.it}
\affiliation{INFN, Sezione di Firenze, Via G. Sansone 1, 50019 Sesto Fiorentino, Italy}
\author{S.~Moretti}\email[E-mail: ]{s.moretti@soton.ac.uk}
\affiliation{School of Physics and Astronomy, University of Southampton, Southampton SO17 1BJ, U.K.}
\affiliation{Particle Physics Department, Rutherford Appleton Laboratory, Chilton, Didcot, Oxon OX11 0QX, UK}
\author{G.~M.~Pruna}\email[E-mail: ]{giovanni-marco.pruna@psi.ch}
\affiliation{Paul Scherrer Institute, CH-5232 Villigen PSI, Switzerland}

\begin{abstract}
\noindent
In this proceeding, we extend a previous analysis concerning the prospects of a future electron-positron collider in testing the 4-Dimensional Composite Higgs Model.
In particular, we introduce two motivated benchmarks and study them in Higgs-Strahlung, for three possible energy stages and different luminosity options of such a machine and confront our results to the expected experimental accuracies in the various accessible Higgs decay channels.
\end{abstract}

\maketitle



\section{Introduction}
\label{Sec:Intro}
\noindent
The discovery at the Large Hadron Collider (LHC) of a Higgs boson~\cite{Aad:2012tfa,Chatrchyan:2012ufa} represents a triumph for the Standard Model (SM). Nowadays, the primary question that requires an answer is whether such a particle belongs to the minimal SM Higgs sector or to some Beyond the SM (BSM) scenario.

Indeed, it is well-known that the SM suffers from the so-called ``hierarchy'' problem (see, {\it e.g.}, \cite{Veltman:1976rt}), pointing out that it could be a low energy effective theory valid only up to some cut-off energy $\Lambda$. Such energy scale is unknown and phenomenological indications about it are missing. Surely though, there is the possibility that $\Lambda$ is lying around the TeV/multi-TeV scale (so that BSM physics could be discovered at the CERN machine in the coming years).

For this reason, many BSM scenarios with new physics at the TeV/multi-TeV scale  were proposed in the last decades. In this spirit, we embrace the possibility that the Higgs particle may be a composite state arising from some strongly interacting dynamics at a high scale instead of being a fundamental state. In this description, the Higgs state arise as a Pseudo Nambu-Goldstone Boson (PNGB) from a particular coset of a global symmetry breaking~\cite{Kaplan:1983fs,Georgi:1984af,Georgi:1984ef,Dugan:1984hq} and it offers an elegant solution for the long-standing hierarchy problem. 

Even in the situation where new physics is outside the discovery range of the present colliders, a composite Higgs state arising as a PNGB has modified couplings with respect to the SM~\cite{Espinosa:2010vn}, hence the measurement of these quantities represents a powerful way to test the possible non-fundamental nature of the newly discovered state. In this case, a TeV/multi-TeV electron-positron collider would represent the cleanest environment for studying possible deviations from the SM signals. For this reason, in this proceeding, we will resume a previous analysis~\cite{Barducci:2013ioa} concerning the potential of the proposed $e^+e^-$ colliders in testing a specific realisation of a composite Higgs model, the so-called 4-Dimensional Composite Higgs Model (4DCHM) of ref.~\cite{DeCurtis:2011yx}, by extending our approach to encompass two new benchmarks and focusing on one of the most interesting Higgs production channel:  Higgs-Strahlung (HS) from $Z$ bosons. As in our earlier paper, we borrow energy and luminosity configurations from machines prototypes such as the International Linear Collider (ILC)~\cite{Behnke:2013xla}, the Compact Linear Collider (CLIC)~\cite{Aicheler:2012bya} and the Triple Large Electron-Positron (TLEP) collider~\cite{Gomez-Ceballos:2013zzn}.


\section{Results}
\noindent
We have implemented the 4DCHM into numerical tools in order to perform dedicated analyses up to event generation. Our simulations have been mainly performed with the CalcHEP package~\cite{Belyaev:2012qa} in which the model had been previously implemented via the LanHEP tool~\cite{Semenov:2010qt}, see~\cite{Barducci:2012kk,Barducci:2013wjc}. Since CalcHEP allows by default the analysis of tree-level processes only, we have also added by hand the one-loop $Hgg$, $H\gamma\gamma$ and $H\gamma Z$ vertices (computed at the leading order without approximations).

For beamstrahlung, CalcHEP implements the Jadach, Skrzypek and Ward expressions of refs.~\cite{Jadach:1988gb,Skrzypek:1990qs}. Regarding the Initial State Radiation (ISR), we adopted the parametrisation specified for the ILC project in~\cite{Behnke:2013xla}, that is:  beam size $(x+y)=645.7$ nm, bunch length $=300$ $\mu$m, bunch population $=2\cdot10^{10}$.

We will be considering throughout three values for the Centre-of-Mass (CM) energy, which are standard benchmark energies for future $e^+e^-$ proto-types: $250$ GeV, $500$ GeV and $1$ TeV. 
Then, we focus on the phenomenology of a Higgs boson obtained via the HS process. 
When combining production cross sections and decay Branching Ratios (BRs), our simulated data will always be related to the experimental accuracies presented in refs.~\cite{Peskin:2012we,Asner:2013psa,Baer:2013cma}. Following their notation, we indicate the production cross section with $\sigma(ZH)$ for HS. In keeping with the aforementioned references, we have will assume a luminosity of $250$/$500$/$1000$ fb$^{-1}$ in correspondence to an energy of $250$/$500$/$1000$ GeV.


In the following subsections we will present several results concerning the studies of the aforementioned Higgs production process, organised as follows. Firstly, we investigate the behaviour of our benchmarks with respect to the mere rescaling of the couplings due to the decoupling of new physics (the so-called ``decoupling limit'').
By considering points that respect exclusion limits from direct and indirect observation of new physics (see~\cite{Barducci:2013ioa} for details on the selection criteria), again we will show that genuine 4DCHM effects cannot be relegated to a simple rescaling of the relevant Higgs couplings, as, for example, the presence of $Z'$ propagator effects in the HS production cannot generally be neglected.

In essence, to generalise our findings, quantitative studies of Higgs boson 
phenomenology in composite Higgs models at future electron-positron colliders should take into account possible effects from realistic mass spectra, whereby extra particles are retained in the calculation of observables, rather than integrated out.

\subsection{Decoupling limit}
\noindent
In order to disentangle rescaling effects (due to both the non-linear realisation of the Goldstone symmetry and the mixing between SM and extra particles) from the ones due to the additional propagators, we have introduced the $R$ and $\Delta$ parameters for inclusive HS production cross section as follows:
\begin{eqnarray}\label{mu_definition}
R=\frac{\sigma(ZH)_{\rm 4DCHM}}{\sigma(ZH)_{\rm SM}}, \qquad \Delta=R-\kappa^2_{HZZ}, \qquad \kappa_{HZZ}=\frac{g_{HZZ}^{\rm 4DCHM}}{g^{\rm SM}_{HZZ}}.
\end{eqnarray}
Then, by numerical computation, we have proven that, if the new class of neutral gauge bosons are completely stripped off the calculations, $\Delta$ tends to $0$ with a negligible deviation $\sim 0.01\%$ related to a slight shift in the $C_V$ and $C_A$ couplings of the SM-like $Z$ to the initial leptons, due to the aforementioned mixing.
Since HS is one of the most useful process to extract information about deviations of the Higgs couplings from the SM values, we are essentially making the generic statement that, even when the CM energy of the collider is below the scale of BSM physics, $f$ in this case (the compositeness scale), where the additional boson and fermion masses of the 4DCHM naturally tend to cluster, the HS cross section is basically always affected by propagator effects.

\begin{figure}[th!]
\includegraphics[width=0.46\textwidth]{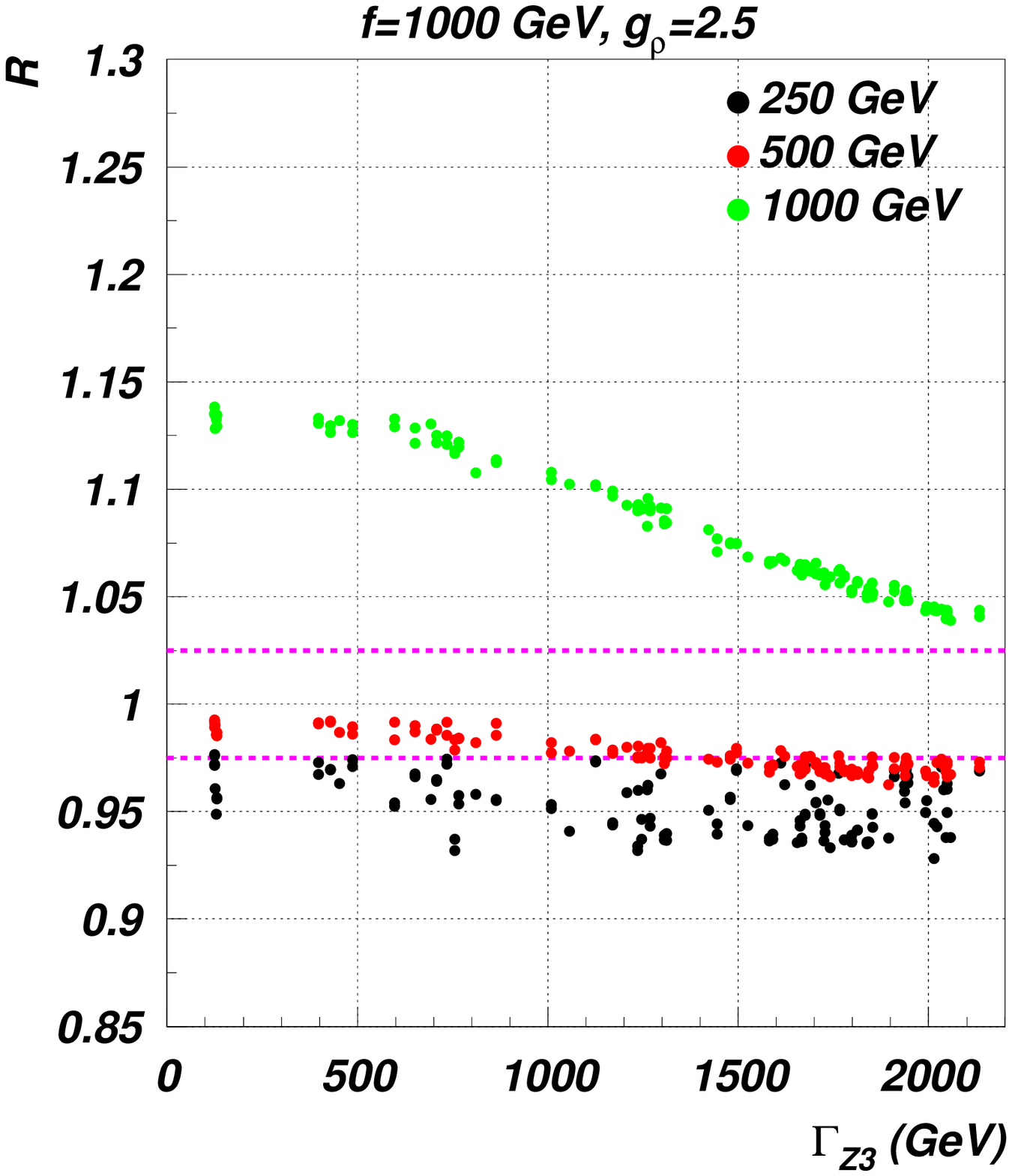}
\includegraphics[width=0.46\textwidth]{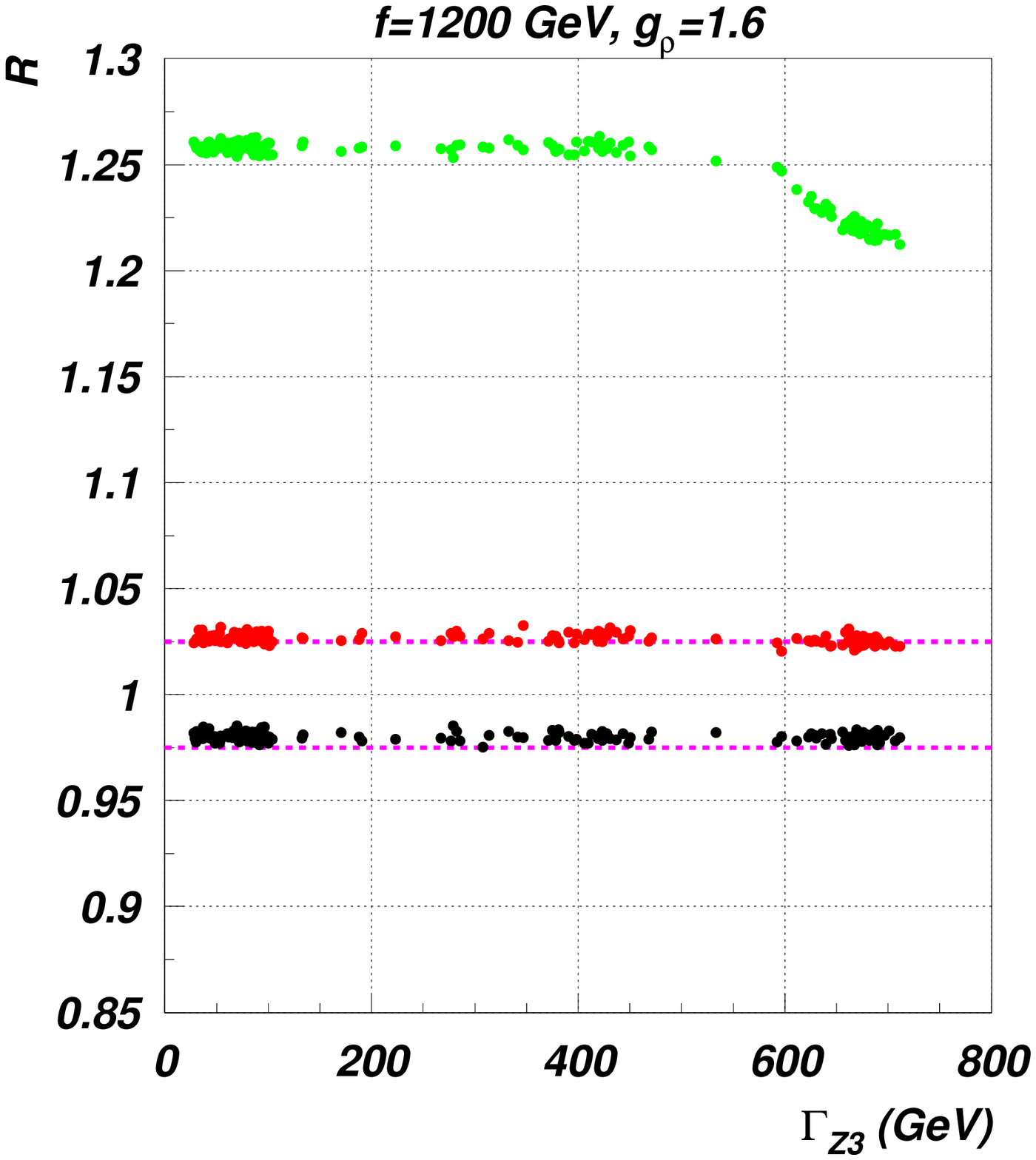}\\
\includegraphics[width=0.46\textwidth]{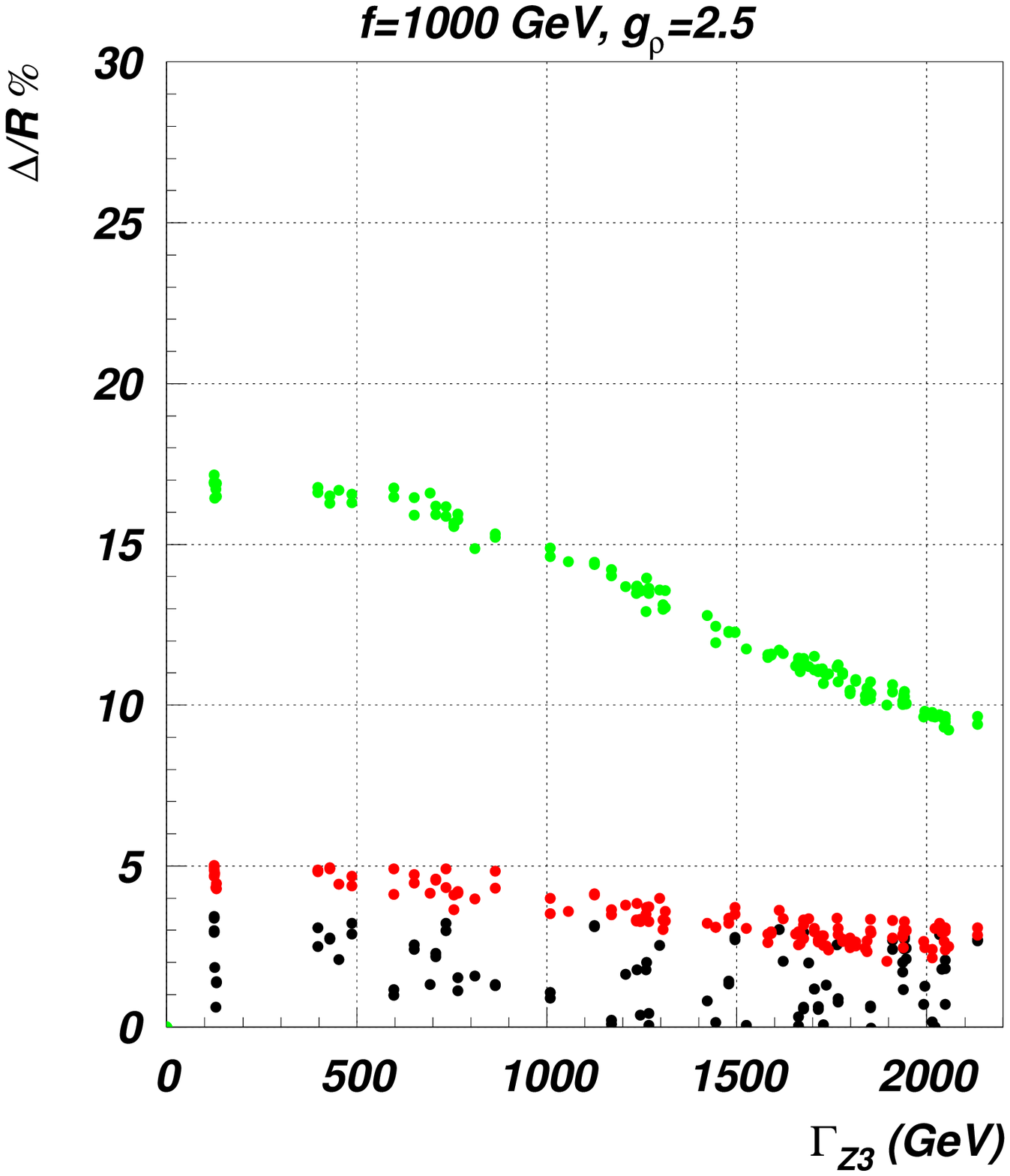}
\includegraphics[width=0.46\textwidth]{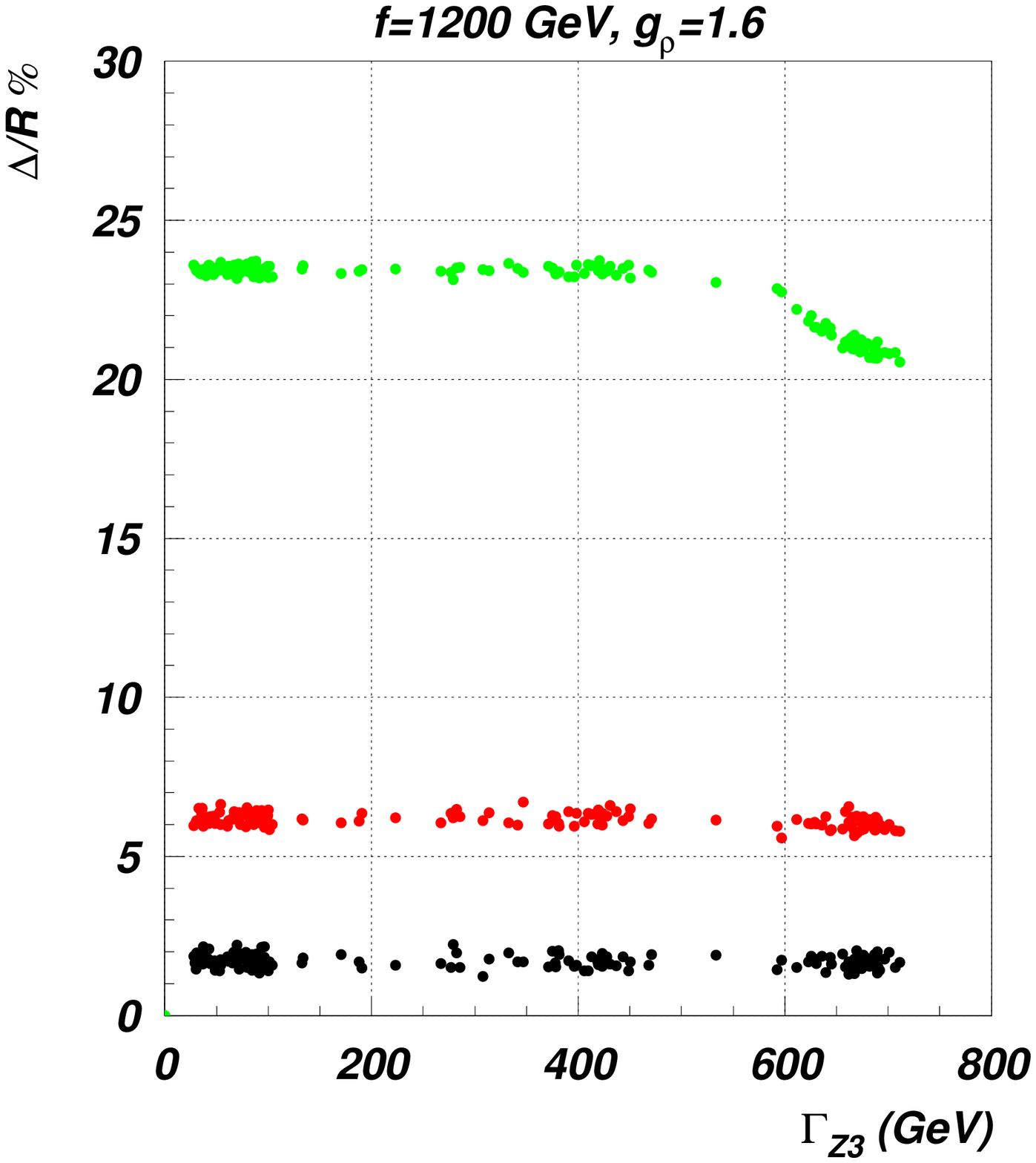}
\caption{The $R$ and $\Delta$ quantities defined in eq.~(\ref{mu_definition}) plotted against the width of the $Z_3$ resonance for the benchmark points with $f=1000$ GeV and $g_\rho=2.5$ (left), $f=1200$ GeV and $g_\rho=1.6$ (right). The dotted purple line represents the experimental precision in determining $R$, according to refs.~\cite{Peskin:2012we,Baer:2013cma}.}
\label{RandN}
\end{figure}

This is well illustrated by fig.~\ref{RandN}, where we quantify the $R$ and $\Delta$ parameters for two new benchmarks ($f=1000~$GeV, $f=1200~$GeV) as a function of the total width of the dominant extra-vectorial contribution, i.e., $\Gamma_{Z_3}$, for the three customary values of CM energy.
The rescaling factors are $\kappa^2_{HZZ}\approx0.94$ for $f=1000~$GeV and $\kappa^2_{HZZ}\approx0.96$ for $f=1200~$GeV.
The slopes present in the plots, the more noticeable the larger the CM energy, show that propagator effects are at work. In fact, the trend of $R$ (or equivalently $\Delta$) is almost constant but, from some threshold ($\sim 600$ GeV) on, it decreases with $\Gamma_{Z_3}$, reflecting the nature of the interference contribution that is proportional to $1/\Gamma_{Z_3}$ when the CM energy is smaller than the $Z_3$ mass involved\footnote{We remark that $M_{Z_3}\sim 2$($2.5$) TeV for $f=1.2$($1$) TeV and $g_\rho=1.6$($2.5$).}. Beside this, the non-zero positive $\Delta$ value definitely points to a constructive interference taking place especially for small values of $\Gamma_{Z_3}$.

As in the previously analysed benchmarks in~\cite{Barducci:2013ioa}, the deviations from the SM limit span from $\sim 2\%$ when $\sqrt{s}=250$ GeV up to $\sim 20\%$ when $\sqrt{s}=1$ TeV. Again, we have verified that the effect is completely due to the constructive interference term arising from the SM-like $Z$ resonance and the $Z_2+Z_3$ contributions, with $Z_3$ being dominant among the two extra vectors. $R$ values are always above the expected ``reduction'' from the decoupling limit: at $\sqrt{s}=1$ TeV and even at $\sqrt{s}=500~$GeV for $f=1200~$GeV the $R$ value is above $1$, which is not compatible with a decoupled scenario.

In fig.~\ref{RandN}, we show that the benchmark with higher values of $M_{Z_{2,3}}\equiv f\times g_{\rho}$ is related to smaller deviations from the decoupling limit, as expected. These results point at the fact that a complete study of composite Higgs models via the HS process should also take into account the possibility of non-decoupled extra resonances.


\subsection{Higgs coupling analysis at $e^+e^-$ colliders in HS}
\label{Sec:VHWW}
\noindent
The presence of extra-vectors in the TeV/multi-TeV range can thus affect the HS cross section due to interference effects. As a consequence, modifications to the various observables can also arise and manifest in the analysis of the Higgs couplings. Therefore, such an alteration would affect the extraction of both the Higgs-vector-vector and vector-fermion-fermion tree-level couplings, as well as the loop-induced couplings $H\gamma Z$, $H\gamma\gamma$ and/or $Hgg$. In other words, these effects can modify the signal strengths in a way that may be detectable with the experimental accuracies expected at future electron-positron colliders.

\begin{figure}[th!]
\includegraphics[width=0.46\textwidth]{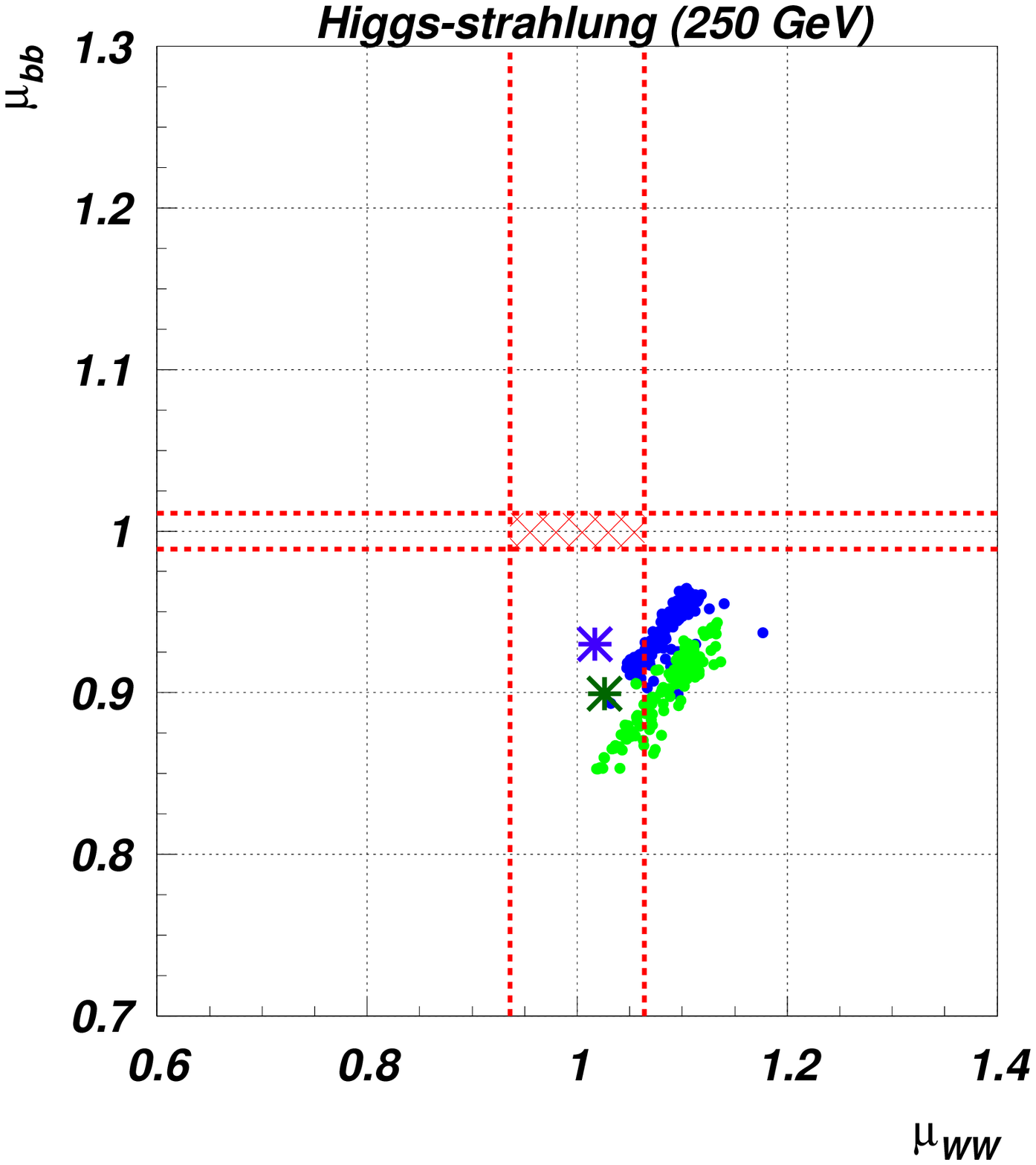}
\includegraphics[width=0.46\textwidth]{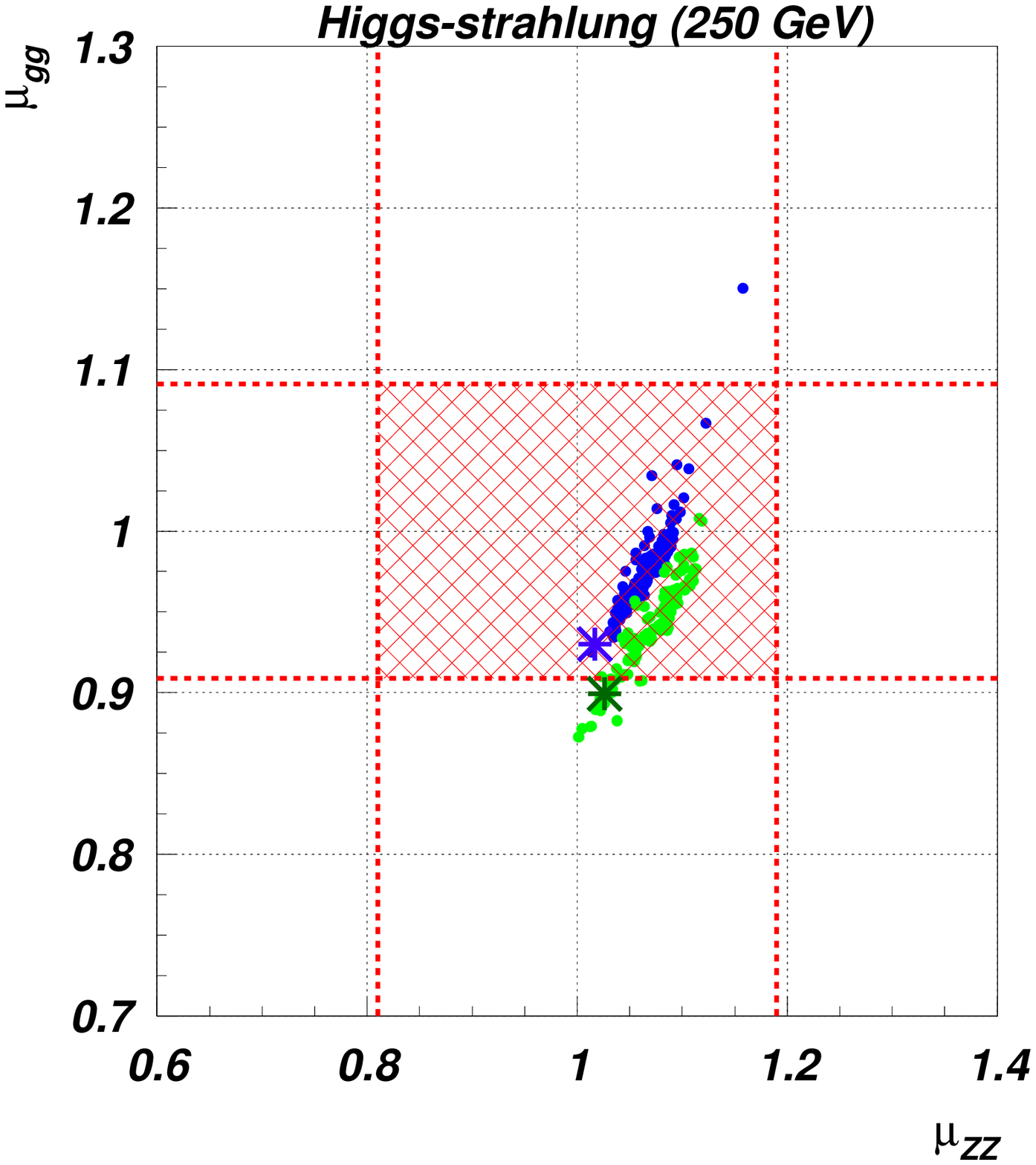}\\
\includegraphics[width=0.46\textwidth]{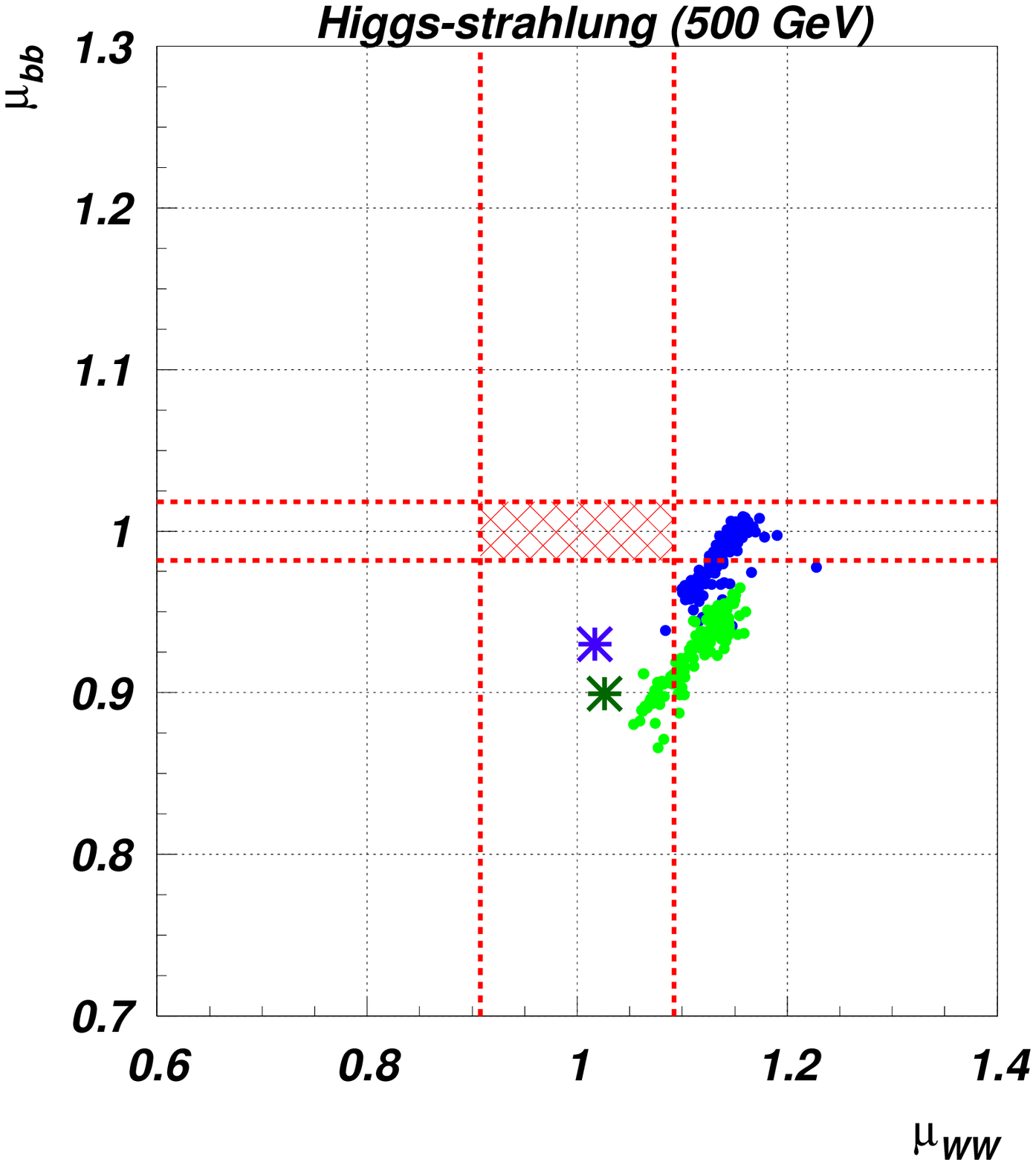}
\includegraphics[width=0.46\textwidth]{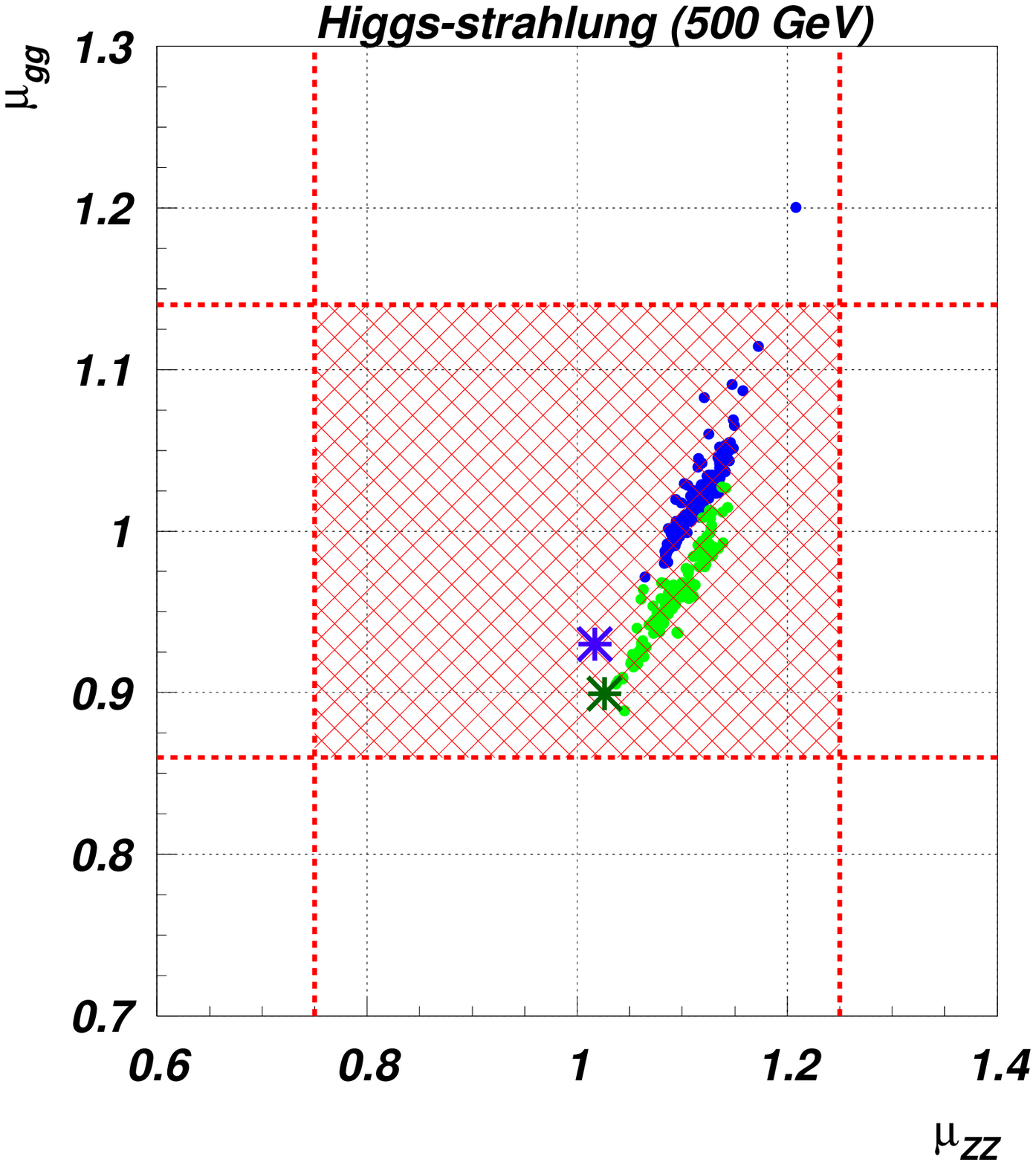}
\caption{Correlations among relevant $\mu_i$ parameters, evaluated at a future $e^+e^-$ collider for two energy and luminosity stages, as detailed in the text, in the HS process. Plots are for two 4DCHM benchmarks, with  $f=1000$ GeV and $g_\rho=2.5$ (green/light-grey points) and $f=1200$ GeV $g_\rho=1.6$ (blue/dark-grey points). The red shadowed area represents the precision limits around the SM expectations according to refs.~\cite{Peskin:2012we,Baer:2013cma}.}
\label{mu}
\end{figure}

In this respect, we present our results in terms of scatter plots for our proposed benchmarks: $f=1000$ GeV, $g_\rho=2.5$ and $f=1200$ GeV, $g_\rho=1.6$. We show the results of these scans in fig.~\ref{mu}, where we notice that the deviations from the case in which the full particle spectrum is not taken into account, represented by the stars, could modify the signal strengths for various channels.

In the case of $\mu_{bb}$ and $\mu_{WW}$, the signal strengths of the $b\bar b$ and $WW$ channels,  
these deviations are fully disentangleable while in the other cases this is not always true, depending on where the scan points fall relative to the SM expectations and according to the corresponding experimental error bars for a particular signature.

Altogether, though, it is clear the potential that future leptonic machines can offer in pinning down the possible composite nature of the Higgs boson discovered at CERN by measuring its ``effective'' couplings to essentially all SM matter and forces.


\section{Conclusions}
\label{Sec:Conclusions}
\noindent
In this proceeding, we extended our previous analysis (see~\cite{Barducci:2013ioa}) to two new benchmarks, albeit
limitedly to the HS channel. We found that, in such concrete realisations of the 4DCHM, the impact of interference effects due to extra TeV/multi-TeV neutral vectors at future $e^+e^-$ colliders is never negligible. We have shown that, as a consequence, also the Higgs signal strengths are affected by such effects. In general, this requires a careful treatment of the methods adopted in the extraction of the SM couplings from the main Higgs production channel, {\it i.e.}, HS, as the real couplings are crucially altered with respect to those emerging in a fully decoupled scenario. This analysis enforces our previous conclusions.


\section*{Acknowledgements} 
\noindent
The work of GMP has been supported by the European Community's Seventh Framework Programme (FP7/2007-2013) under grant agreement n.~290605 (PSI-FELLOW/CO\-FUND). DB and SM are financed in part through the NExT Institute. GMP would like to thank the ECT* in Trento for hospitality while part of this work was carried out.


\end{document}